\documentclass[epj]{svjour}
\usepackage{graphicx}
\input{epsf}
\usepackage{amssymb}
\begin{document}

\title{Kolmogorov turbulence, Anderson localization and \\
KAM integrability} 

\author{D.L.Shepelyansky \inst{1,2} }
\institute{Laboratoire de Physique Th\'eorique du CNRS, IRSAMC,  
  Universit\'e  de Toulouse, UPS, 31062 Toulouse, France
\and
http://www.quantware.ups-tlse.fr/dima}

\date{March 4, 2012}



\abstract{
The conditions for
emergence  of Kolmogorov turbulence,
and related weak wave turbulence, in finite size systems
are analyzed by analytical methods
and numerical simulations of simple models.
The analogy between Kolmogorov energy flow
from large to small spacial scales
and conductivity in disordered
solid state systems is proposed.
It is argued that the Anderson localization
can stop such an energy flow.
The effects of nonlinear wave interactions
on such a localization are analyzed.
The results obtained for finite size system models
show the existence of an effective
chaos border between the Kolmogorov-Arnold-Moser (KAM)
integrability at weak nonlinearity, when
energy does not flow to small scales,
and developed chaos regime 
emerging above this border
with the Kolmogorov turbulent energy flow
from large to small scales.
}

\maketitle

\section{ Introduction}
The concept of Kolmogorov turbulence 1941 \cite{kolm41,obukhov}
describes how the energy flows from large to small space
scales in a turbulent regime. According to this concept
an energy is injected on large scales, e.g. by wind,
and it is absorbed on small scales due to dissipation.
As a result a stationary algebraic distribution $\epsilon_k \propto k^{-5/3}$
of energy flow is established over wave modes $k$  \cite{kolm41,obukhov}.
This concept was shown to be generic not only for hydrodynamic
turbulence but also for other types of nonlinear waves.
This phenomenon became
known as the weak turbulence. With the help of diagrammatic technique
Zakharov and Filonenko 1967 \cite{filonenko}
derived the kinetic equation for weak turbulence of capillary waves
and demonstrated the existence of stationary
algebraic energy flows similar to those of Kolmogorov \cite{kolm41}. 
Later the concept of weak wave turbulence was
generalized for various types of nonlinear waves
as it is described in detail in 
\cite{zakharovbook,nazarenkobook}.
However, an enigma of turbulence still remains as it is 
clearly stated in a poetic claim \cite{zhakharov1992}
{\it ``Through mechanisms still only partially understood,
wind transfers energy and momentum to surface water waves.''}

Indeed, the kinetic equation for energy flow from small to large
wave vectors is derived in the regime of weak nonlinearity
and random phase approximation, known in plasma as
a quasilinear approximation \cite{sagdeev}.
This hypothesis is at the basis of the whole theory
as it is directly stated by Zhakharov and Filonenko
at the first paragraph of their fundamental paper \cite{filonenko}:
{\it ``In the theory of weak turbulence nonlinearity of waves is assumed to be
small; this enables us, using the hypothesis of the random nature of 
the phase of individual waves, to obtain the kinetic equation 
for the mean square of the wave amplitudes''}. However, 
in finite size systems one has a 
a discrete spectrum of linear modes
and the dynamical origins and conditions for validity
of the random phase approximation and hence, for validity of
the Kolmogorov concept of energy flow from small to large space scales,
are still to be established.

In fact, it is known that a weak nonlinearity does
not always lead to global ergodicity and chaos over all wave vectors
or over all degrees of freedom.
The well known example is the Fermi-Pasta-Ulam (FPU) problem \cite{fpu}
where nonlinearity should be strong enough
to generate global chaos as it was pointed by
Chirikov and Izrailev \cite{chirikovfpu}.
At present the FPU problem  still remains under active
studies aiming to understand its ergodic properties
in the limit of low energy or weak nonlinearity
\cite{gallavotti,benettin}. For the FPU problem it was shown  
that, in  the limit of very small nonlinearity combined with
the resonant approximation,
which is typically used for the derivation
of the kinetic equation for wave \cite{filonenko,zakharovbook,nazarenkobook},
it is possible to have chaos only at small $k-$vectors
with exponential decay of energies at high $k-$vectors
(see \cite{dlsfpu} and Fig.2 therein). Thus this result
\cite{dlsfpu} shows that  certain conditions 
are required for the emergence of algebraic stationary flows
in the weak turbulence in finite systems. 

Indeed, in the case of a few degrees of freedom it is established
that in the limit of weak nonlinear perturbation almost all
phase space of a typical Hamiltonian nonlinear system remains in an
integrable regime known as the Kolmogorov-Arnold-Moser (KAM)
integrability (see e.g. \cite{chirikov1979,lichtenberg} and Refs. therein).
The rigorous form of this statement is known as the KAM theory.
The transition to global chaos and ergodicity
requires that nonlinearity exceeds a chaos border
which can be determined by numerical simulations or, in a number of cases, 
analytically by the Chirikov criterion 
\cite{chirikov1979,lichtenberg},\cite{dlschirikov}.
Below the chaos border a main part of the phase space remains integrable
and a chaotic spreading is possible only due to the Arnold diffusion
via tiny chaotic web in a separatrix vicinity   
 \cite{chirikov1979,lichtenberg},\cite{chirikovvech}.
In the limit of small nonlinear perturbation the measure
of these chaotic layers drops exponentially
\cite{chirikov1979,lichtenberg} even if for a larger number of
degrees of freedom this exponential regime
can become valid only at very very small perturbations
\cite{chirikovvech,mulansky2011}.

Of course, the initial concept of Kolmogorov turbulence
and weak wave turbulence
considers systems of large scale with a continuous
spectrum of linear waves \cite{zakharovbook,nazarenkobook}.
However, the modern experiments on wave turbulence
are done with systems of finite size 
(see e.g. \cite{lukaschuk,water,maurel}).
Also all numerical simulations
are done with the finite size systems
(see e.g. 
\cite{nazarenkolvov,nazarenko2006},\cite{kartashova2007},
\cite{kartashova2008pre,nazarenko2009},
\cite{kartashova2009,tsubota2011}).
It is recognized that 
in such finite systems certain resonant conditions
play important role for energy flows in $k-$space
\cite{nazarenkolvov,kartashova2007},
\cite{kartashova2008pre,nazarenko2009},
\cite{kartashova2009}.
Indeed, it is natural to assume that
a discreetness of linear frequencies in a finite
system can stop chaotic spreading
if nonlinear frequency shifts become smaller
than a typical spacing between linear resonant modes.
Such a criterion has been put forward in \cite{dlsfpu}
where its validity was confirmed for the FPU model 
in the resonant approximation.

However, the situation may be even more complicated
and the energy spreading can be suppressed even
in systems with everywhere dense spectrum.
Indeed, in disordered systems it is known that
the spreading of quantum (linear) waves can be 
stopped due to the phenomenon of  Anderson localization
\cite{anderson1958,montambaux} even if the spectrum
of linear waves is everywhere dense and
classical particles can diffusively spread over the whole system.
The effects of weak nonlinearity on the Anderson
localization are now under active investigations
of different groups (see e.g. 
\cite{dls1993,dlsphysd},
\cite{molina,pikoshep},
\cite{tietpik,wang},
\cite{garciamata,flachkrimer},
\cite{laptyeva,mulpik},
\cite{johansson,pikfishman}). Usually the numerical 
simulations are done for  
a discrete Anderson nonlinear Shr\"oringer equation (DANSE),
which allows to perform numerical simulations in an efficient way
up to very large times and thus to study
wave packet spreading in space (see e.g. \cite{pikoshep}).
It is established that at moderate nonlinearity $\beta$
a subdiffusive wave packet spreading  continues up to enormously large times
being at least by $8$ orders of magnitude larger than a typical time
scale in a system \cite{dls1993,pikoshep},
\cite{garciamata,flachkrimer},
\cite{mulpik}. 
However, at small nonlinearity being below a certain border $\beta_c$
($\beta < \beta_c$) the spreading
is  absent up to maximal numerically available times.
It should be pointed out that the
exact mathematical results are difficult to obtain
even in the limit of $\beta \rightarrow 0$ since the spectrum of 
the linear problem is everywhere dense 
so that resonances appear on large space scales \cite{wang}.
Due to that it is difficult to develop the mathematical KAM theory
in such a regime.

Below the critical value of nonlinearity $\beta < \beta_c$
the wave amplitudes decay exponentially inside a disordered
layer \cite{dlsphysd,pikoshep},
\cite{tietpik,johansson} in a way
similar to a disordered linear media
in the regime of  the Anderson insulator characterized 
by an absence of conductivity in such a system \cite{montambaux}.   

It is rather appealing to expect that the spreading in $k-$space
of Kolmogorov turbulence will go in a way similar to the DANSE case.
In fact, for the quantum Chirikov standard map, 
known also as the kicker rotator,  
it is established that for a periodic driving in time
the quantum localization of dynamical chaos
takes place in the momentum $k-$space in a way
similar to the Anderson localization in a coordinate space
\cite{kr1,fishman1982},
\cite{dlskr}. The extension of this linear wave model
to the case of the kicked nonlinear Schr\"odinger equation (KINSE) 
was proposed in \cite{kicknse}. In this KINSE model 
 the nonlinear wave interaction
takes place locally in space while we are interested
in the energy spreading in the momentum $k-$space
as it is usually the case for the weak wave turbulence \cite{zakharovbook}.
In this respect the situation is different compared to
the DANSE model where both nonlinear wave interaction and
spreading take place in coordinate space \cite{pikoshep}.
Even if it was argued that there is a certain similarity
between these two cases \cite{dls1993} a special more detailed
analysis of the KINSE model is required.
In this work the KINSE model is studied on a large time scales and 
the links with the DANSE model are traced in a firmer way.
The implications of the obtained results for the Kolmogorov turbulence in
finite size systems are discussed.

\section{An example from the FPU problem}

Let us discuss briefly the effects of discreteness
of linear wave spectrum on example of the $\alpha-$FPU problem
following the results presented in \cite{dlsfpu}.
It is shown there that in the long wave limit
the system dynamics can be described by an effective
renormalized Hamiltonian $H_{RN}$ (see Eq.(4) in 
\cite{dlsfpu}). This Hamiltonian is only
a resonance approximate description
of the initial FPU problem.
However, it is important to see 
what are the properties of this 
Hamiltonian itself,
since it has a typical form of 
resonant Hamiltonians considered
in the theory of weak turbulence.
This Hamiltonian $H_{RN}=H_{RN0}+H_{RNint}$ has
an unperturbed part $H_{RN0} \propto k^3 J_k$
corresponding to the renormalized linear spectrum
of long waves and a part describing the 
renormalized resonant interacting waves $H_{RNint} \propto
\cos(\phi_{k_2+k_1}-\phi_{k_2}-\phi_{k_1})$
usually used in the theory of weak turbulence
\cite{zakharovbook} (here $(J_k,\phi_k$ 
are conjugated pairs of action-phase variables). 
It is shown in \cite{dlsfpu}
that even if the dynamics of Hamiltonian $H_{RN}$
is chaotic for waves with $k \sim 1$ this chaos does not 
create energy flow to high wave vectors $k$ and the 
energy density drops exponentially at large $k$
(see Fig.2 in \cite{dlsfpu}).
This shows that the random phase approximation
assumed in the weak turbulence theory
can be not correct 
and that there can be no Kolmogorov flow from
small to large wave vectors in finite size systems.

\section{The KINSE model description}

We focus here on the KINSE model described by
\begin{equation}
\begin{array}{c}
i \hbar{{\partial {\psi}}/{\partial {\tau}}}  =
 - {{\partial^2 {\psi}}/2{\partial^2 {x}}} 
+{\beta} |\psi|^2 \psi \\
 - k \cos x \; \psi \sum_{m=-\infty}^{\infty} \delta(\tau-mT) \; ,
\end{array}
\label{eq1}
\end{equation}
where 
$\beta$ characterizes nonlinearity,
$k$ is kick potential amplitude, $T$ is a period between 
$\delta-$function kicks
periodic in time, in the following we put $\hbar=1$.
We impose the periodic boundary conditions
with $\psi(x+2\pi)=\psi (x)$ and normalization condition
$\int_0^{2\pi} |\psi(x)|^2 dx =1$.
The linear wave expansion has the form
$\psi(x)=  \sum_n \psi_n \exp(-inx)/\sqrt{2\pi}$
with conserved normalization $\sum_n|\psi_n|^2=1$.
The second moment is defined as
$\sigma(t)=\sum (n-n_0)^2 |\psi_n(t)|^2$,
in the following we measure time $t$ in the number of kicks
($n_0$ is the initial mode).

The linear model at $\beta=0$ has been studied in a great detail
(see e.g. \cite{kr1,fishman1982},\cite{dlskr}).
The semiclassical regime of this model corresponds
to $k \gg 1$, $T \ll 1$ with the classical chaos parameter $K=kT =const$.
In fact $K$ is the chaos parameter in the Chirikov standard map
which describes the classical dynamics \cite{chirikov1979,lichtenberg}.
The classical dynamics is globally chaotic for
$K >1$ with a diffusive growth on energy
proportional to $\sigma$ with $\sigma = D t$
and the diffusion rate $D \approx k^2/2$ for $K > 4$.
In the quantum case this diffusion is localized due to 
quantum interference effects
with exponentially localized Floquet states 
$|\psi_n| \propto \exp(-2|n-n_0|/\ell)$
and the localization length $\ell \approx D/2$ \cite{kr1,dlskr}.
This dynamical, or Chirikov localization, is similar to the Anderson 
localization in disordered linear lattices where the momentum
states $n$ play the role of spacial coordinate \cite{fishman1982,dlskr}.
In a difference from the Anderson localization,
which takes place in presence of disorder,
the Chirikov localization takes place in a purely
dynamical system without any randomness, but due to dynamical chaos
the mechanism of localization of dynamical diffusion
is similar to the one of Anderson localization.
In absence of kick $(K=0)$ the model is reduced to the
integrable nonlinear Sch\"odinger equation.
The KINSE model has been realized experimentally with the
cold atoms and Bose-Einstein condensates (BEC)
in kicked optical lattices (see e.g. \cite{raizen,hoogerland}).
The Chirikov localization at $\beta=0$ was observed
experimentally \cite{raizen}, the studies of effects of BEC nonlinearity
on this localization are now within 
experimental reach \cite{hoogerland}.

The KINSE model was introduced and studied 
in \cite{kicknse}. It was shown there that a narrow soliton
has a long live time during which it
follows an integrable or chaotic trajectory
of the Chirikov standard map.
In a regime when a soliton is destroyed
it was found that there is still a suppression
of classical diffusive growth of $\sigma$. 
In \cite{dls1993} it was conjectured that 
this growth is similar to the case
of the nonlinear kicked rotator (KNR) model,
where there is a nonlinear phase shift 
for linear modes in momentum  representation
(it takes place during each kick period
and is proportional to $\beta |\psi_n|^2$).
The evolution of the KNR model is described by a nonlinear map 
for the wave function \cite{dls1993}:
\begin{equation}
\bar{{\psi}}_{n} = \exp(-iT \hat{n}^2/2-i\beta |\psi_n|^2) 
\exp(-ik\cos \hat{x}) \psi_n \; .
\label{eq2}
\end{equation}
Here the bar marks the wave function after one 
period of perturbation, the operator $\cos \hat{x}$
gives the Bessel coupling between the momentum states $\psi_n$.
It  was also argued  \cite{dls1993} that the KNR 
behavior is similar to the case of the DANSE model,
where the linear modes are also 
exponentially localized. 
The DANSE has the form
\begin{equation}
i {{\partial {\psi}_{n}}/ {\partial {\tau}}}
=E_{n}{\psi}_{n}
+{\beta}{\mid{\psi_{n}}\mid}^2 \psi_{n}
 +V ({\psi_{n+1}}+ {\psi_{n-1})}\;,
\label{eq3}
\end{equation}
where 
$\beta$ characterizes nonlinearity,
$V=1$ is a hopping matrix element, on-site
energies are randomly distributed in the range
$-W/2 < E_n < W/2$. The spreading in this model
at moderate $\beta \sim 1$
is characterized by a subdiffusive growth
\begin{equation}
\sigma(t) \propto t^{\alpha}
\label{eq4}
\end{equation}
with the exponent $\alpha \approx 0.3 - 0.4$
(see details in \cite{pikoshep,garciamata},\cite{flachkrimer}).
The similar values of the spreading exponent $\alpha$
have been found for the KNR model 
\cite{dls1993,garciamata},\cite{flachnkr}.

If to consider the lattice sites
$n$ in (\ref{eq3}) as the 
momentum states then it is clear that
the nonlinear coupling in (\ref{eq3}) 
is local while in the KINSE model (\ref{eq1})
it is strongly nonlocal in momentum or $k-$space,
as it is usual for nonlinear 
wave interaction in the regime of weak turbulence.
Hence, the KINSE model is more adapted to the studies
of the Kolmogorov turbulence in finite systems.
Indeed, the kicks take place on a spacial scale of the whole
system (of size $2\pi$) and can be considered as
a model of wind which pumps energy from small $k-$vectors
(small $|n|$) to large ones. 
The numerical studies of the model (\ref{eq1})
are presented in the next Section.

\section{Numerical results for KINSE model}

To integrate numerically the evolution described by (\ref{eq1})
it is convenient to use unitary small step integrator
making small kicks in coordinate space with the local 
space nonlinearity and
returning back and forth to the momentum space $n$
with the fast Fourier transform as described in \cite{kicknse}.
However, on very large times $t \sim 10^6$
the nonlinearity can generate exponential instability
on high $k-$modes which is of a purely numerical origin
related to discretization. To eliminate this artificial
instability the simulations are done on each step in an enlarged
space size (approximately 4 times larger than the 
size of physical modes) and after each small step
the amplitudes out of the physical size
are suppressed to zero. Such a method is similar
to the aliasing approach \cite{aliasing}
which efficiently suppresses numerical instability on high modes
\cite{athanks}.
Usually the simulations are done
with the total number of states $N_{tot} =2^{10}$
and the size of physical state $N=220$ ($-110 \leq n \leq 110$).
The number of small steps $N_s$ on one kick period varied between $100$
and $1000$ with a special check that it does not affect
the accuracy of the results. The numerical integration method
preserves the total probability up to a numerical double precision.
The maximal value of $t$ reached in the numerical simulations is
$t=10^7$.

\begin{figure}[ht]
\begin{center}
\includegraphics[width=0.41\textwidth]{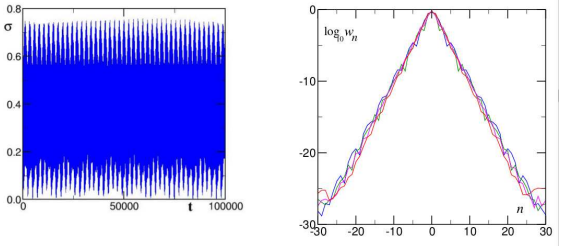}
\end{center}
\vglue 0.2cm
\caption{Left panel: dependence of the second moment $\sigma$
on time $t$ measured in number of kicks in (\ref{eq1}).
Right panel:  probability distribution $w_n$
over linear wave modes $n$ at
times $t=10^3$ (blue), $10^4$ (green), $10^5$ (magenta),
$10^6$ (red) (all curves are superimposed).
Here $\beta=1$, $T=2$, $k=0.3$, $K=kT=0.6$
the initial state is at zero mode $n=0$.
\label{fig1}}
\end{figure}

The time evolution of the second moment $\sigma(t)$
and probability distribution over linear modes
$w_n=|\psi_n|^2$ in (\ref{eq1}) 
are shown in Fig.~\ref{fig1} for the case of
moderate nonlinearity $\beta=1$ and
a small kick amplitude $k=0.3$
corresponding to the chaos parameter
$K=0.6$ being below the global chaos border
$K_c \approx 1$ for the Chirikov standard map
\cite{chirikov1979}. It is clear that there are only
quasi-periodic oscillation of $\sigma$ and that
the distribution in momentum space
remains exponentially localized.
Thus, in this regime there is no energy
flow to small scales and random phase approximation
assumed in weak turbulence \cite{filonenko,zakharovbook},
\cite{nazarenkobook}
is not valid. This result is similar to 
a usual observation that a small wind (small $k$ here)
is not able to produce a turbulent storm.
\begin{figure}[ht]
\begin{center}  
\includegraphics[width=0.41\textwidth]{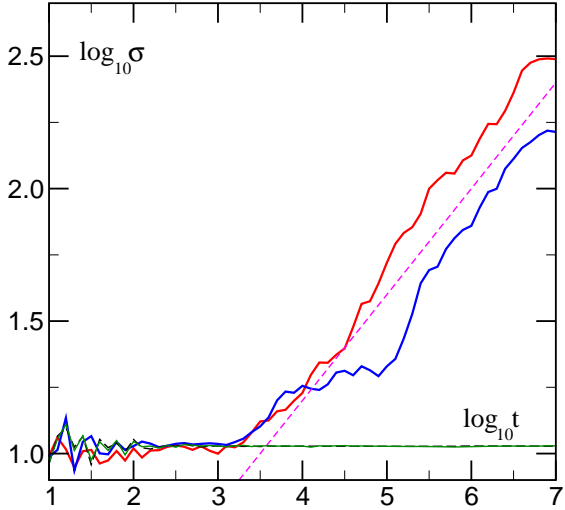} 
\vglue 0.3cm
\caption{Dependence of the second moment 
$\sigma$ of probability distribution over 
wave modes $n$ on time $t$ for (\ref{eq1}).
Here, $k=3$, $T=2$, $K=kT=6$, $\beta=0.5$ (blue curve)
and $\beta=1$ (red curve); $\sigma(t)$ is averaged over 
logarithmically equidistant time intervals;
the initial state is $n_0=0$.
The dashed line shows an anomalous diffusion
$\sigma \sim t^\alpha$ with the exponent $\alpha=0.4$;
the fit of data in the range $3.5 \leq \log_{10} t \leq 7$
gives $\alpha =0.346 \pm 0.014$ (for $\beta=0.5$),
$\alpha =0.438 \pm 0.007$ (for $\beta=1$).
The horizontal dashed black curve shows data
at $\beta=0$, it practically coincides with
the data for $\beta=0.05$ shown by green curve.
\label{fig2}}
\end{center}
\end{figure}

Let us now consider the regime above the classical chaos 
border with $K=6$. In his regime the classical system,
described by the Chirikov standard map \cite{chirikov1979},
has diffusive energy growth with 
$\sigma \approx k^2 t/2 \approx 4 t$.
The results of Fig.~\ref{fig2}
at $\beta=0$ show that
this diffusion is localized by quantum interference
with the localization length $\ell \approx k^2/4 \approx 2.2$
and $\sigma \sim \ell^2 \sim 10$ (the value of sigma is
slightly higher than $\ell^2$ due to mesoscopic fluctuations
of $\ell$ \cite{dlskr}). A very weak nonlinearity
$\beta=0.05$ does not affect this localization
which persists up to maximal times $t=10^7$ reached
in numerical simulations. We note that following 
\cite{pikoshep} the values of $\sigma$ are averaged over
logarithmically equidistant time intervals that
suppress fluctuations at large times. 

However, at moderate nonlinearity $\beta=0.5$ and $\beta=1$
the second moment shows a subdiffusive growth
with the algebraic exponent $\alpha \approx 0.4$.
The exact fit values of $\alpha$ 
are given in the caption of Fig.~\ref{fig2}.
As in \cite{pikoshep} the statistical error
bars are relatively small but there are long
time correlations that can affect the real
$\alpha$ value on large time scales
as discussed in \cite{pikoshep,garciamata},
\cite{flachkrimer,laptyeva}.
A similar value of $\alpha \approx 0.4$
has been found by A.S.Pikovsly for numerical simulations of the KINSE
model by another numerical method for $t\leq 10^6$ \cite{pikoprive}.  

\begin{figure}[ht]
\begin{center}
\includegraphics[width=0.22\textwidth]{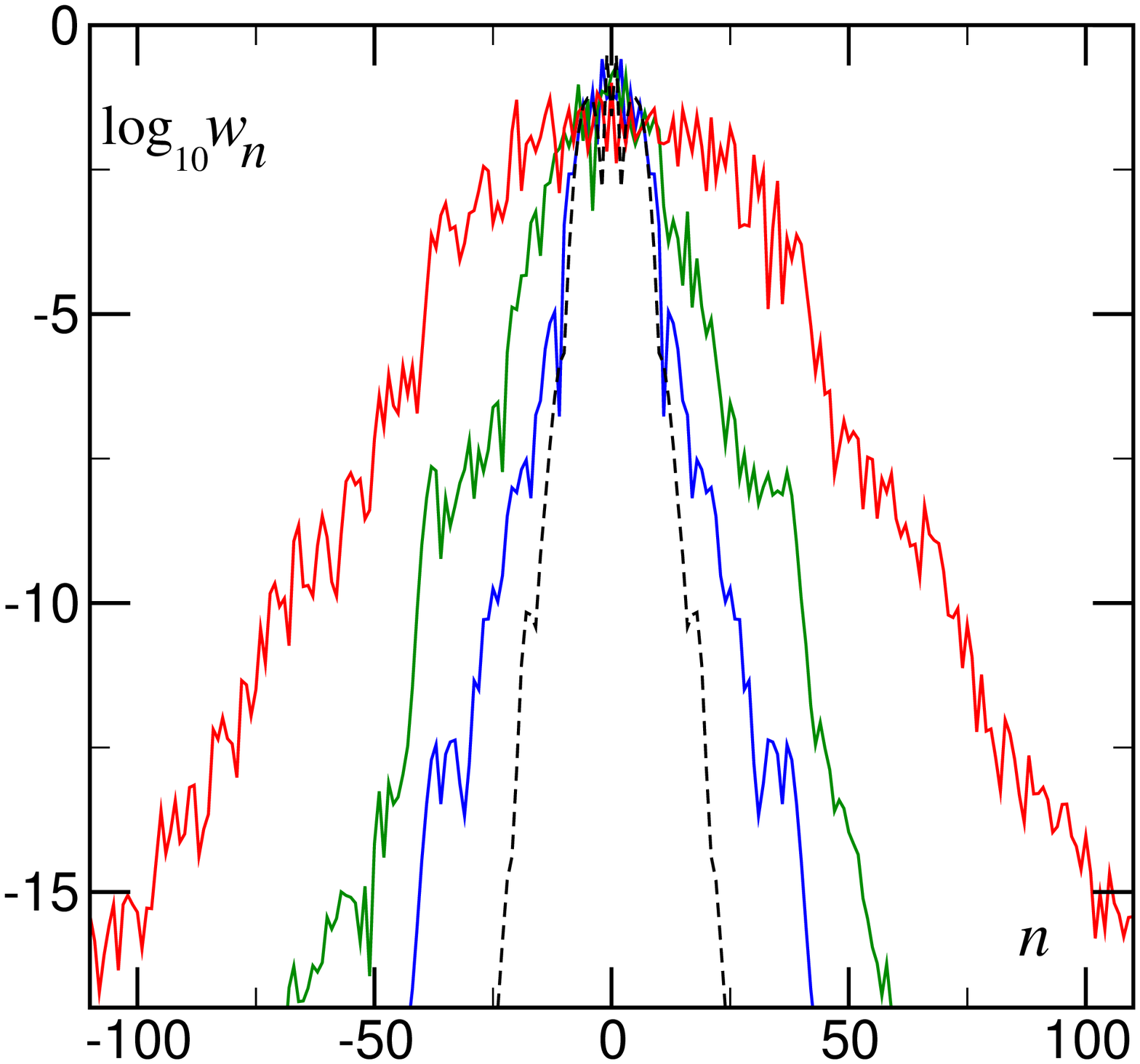}
\hglue 0.5cm
\includegraphics[width=0.22\textwidth]{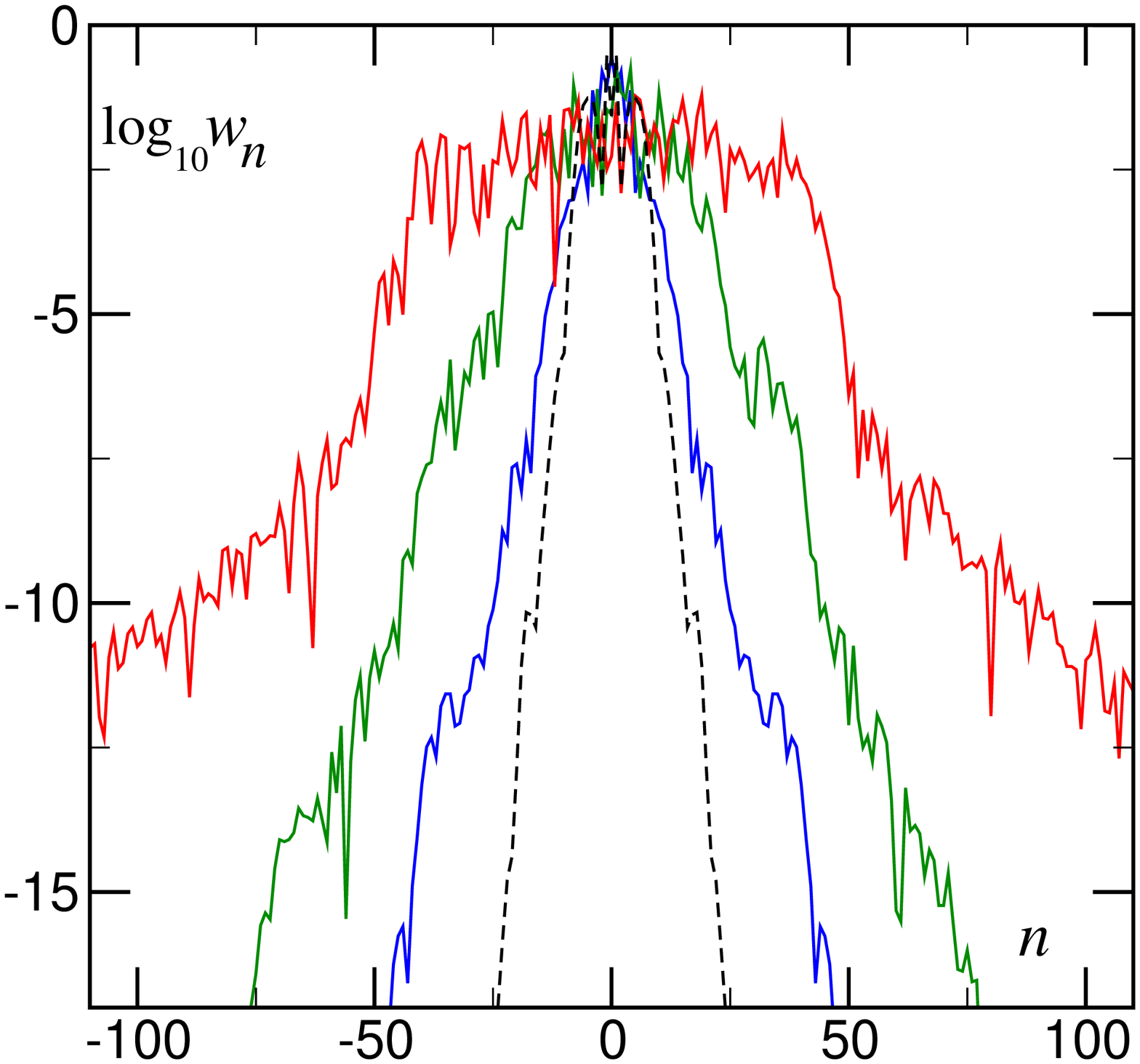}
\end{center}
\vglue 0.2cm
\caption{Probability distribution $w_n$
over linear wave modes $n$ in (\ref{eq1}) at
times $t=10^3$ (blue), $10^5$ (green), $10^7$ (red)
for $\beta=0.5$ (left panel), $1$ (right panel);
other parameters are as in Fig.~\ref{fig2}.
The dashed curve shows the probability distribution
at $\beta=0$, $t=10^7$.
\label{fig3}}
\end{figure}

\begin{figure}[ht]
\begin{center}
\includegraphics[width=0.41\textwidth]{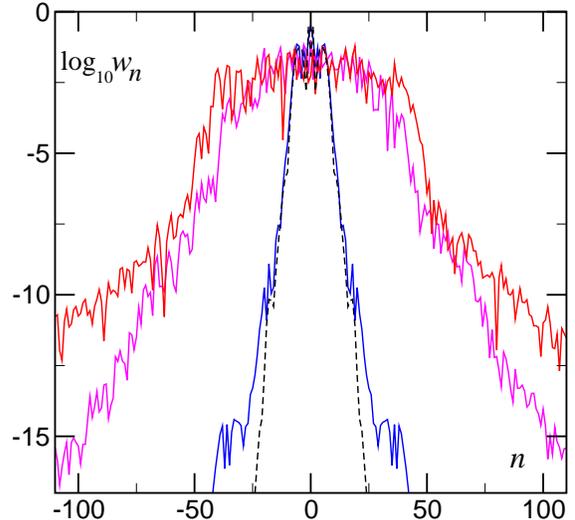}
\end{center}
\vglue 0.2cm
\caption{Probability distribution $w_n$
over linear wave modes $n$ in (\ref{eq1}) at
time $10^7$ 
for $\beta=0.0$ (dashed black), $0.05$ (blue), 
$0.5$ (magenta),
$1$ (red);
other parameters are as in Fig.~\ref{fig2}.
\label{fig4}}
\end{figure}

The evolution of probability distribution $w_n$ with time
is shown in Fig.~\ref{fig3} for $\beta=0.5; 1$.
It is clear that nonlinearity destroys localization:
an approximately flat plateau of probability is formed
(``chapeau'') which size is slowly growing with time.
The data for $w_n$ at large time $t=10^7$
show that there is a significant increase
of the distribution size for $\beta=0.5, 1$,
while for $\beta=0.05$ the probability remains
localized in a way similar to the linear case $\beta=0$.
These data qualitatively confirm 
existence of chaos border in nonlinearity
with a certain $\beta_c \sim 1/10$. However,
it is not excluded that some very slow processes related to
the Arnold diffusion can lead to ``escape'' of some small probability
to larger $n$ values at exponentially large times.

The results obtained for the KINSE model (\ref{eq1})
show that the behavior of this model
is qualitatively similar to the one found in 
the  KNR (\ref{eq2}) and DANSE  (\ref{eq3}) models:
the spreading in the momentum remains localized
below a certain chaos border in nonlinearity strength,
while  above this border the spreading continues in
a subdiffusive way with the algebraic exponent
being close to the value $\alpha \approx 0.4$ 
found previously for KNR and  DANSE.
The main difference 
from DANSE is that for KINSE we have interchange between
momentum and coordinate spaces. However, such a similarity
between momentum  and coordinate space is 
well known \cite{fishman1982,dlskr}
and was already discussed for the KNR model in \cite{dls1993}.
The main new aspect of the KINSE model is a long range
coupling in the momentum space due to the locality of 
wave interaction in the coordinate space. However, 
due to a localized nature of linear
eigenmodes in the momentum space this long range
coupling still can give transitions only 
on a size of localization length $\ell$
and this does not produce a qualitative
difference from the case of short range 
interation appearing in DANSE.
A similar situation appears for the model
of two particles with Coulomb interaction
in the regime of Anderson localization
as it is discussed in \cite{tip}.

\section{Kolmogorov turbulence in Sinai billiard}

On a first glace one can get an impression that the KINSE
model is a rather specific one: it is one-dimensional,
there are kicks etc. However, it is known that
the Chirikov standard map is generic and
describes a variety of real physical systems 
\cite{chirikov1979,lichtenberg}. Also it is known
that the Chirikov localization found first
in the linear KINSE and NKR models at $\beta=0$
appears in a variety of real systems
(see e.g. \cite{fishmanschol,dlsschol}).
Thus we expect that the behavior found here for 
the KINSE model 
at finite nonlenearity $\beta$ 
will appear in more realistic systems.

\begin{figure}[ht]
\begin{center}
\includegraphics[width=0.41\textwidth]{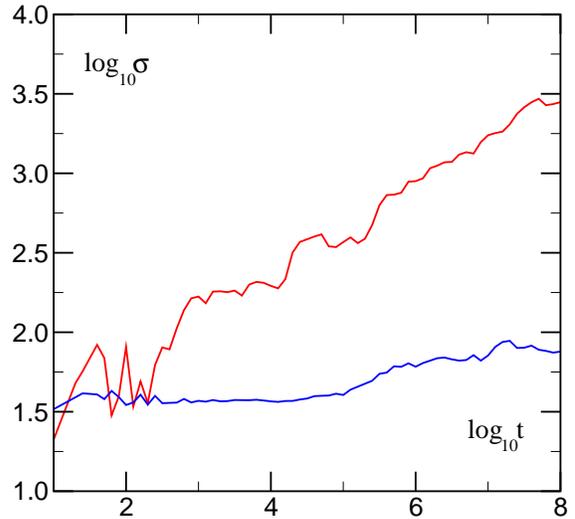}
\end{center}
\vglue 0.2cm
\caption{Dependence of the second moment $\sigma$
in the DANSE model (\ref{eq3}) at
$W=4$, $V=1$, $\beta=1$ (red curve)
and in the DANSE model with additional static 
field potential (SDANSBIL model)) with $\delta E_n = f |n|$
at $f=0.5$ (blue curve) at the above parameters and 
the same disorder realization. The initial state is
at $n=0$.
\label{fig5}}
\end{figure}

\begin{figure}[ht]
\begin{center}
\includegraphics[width=0.22\textwidth]{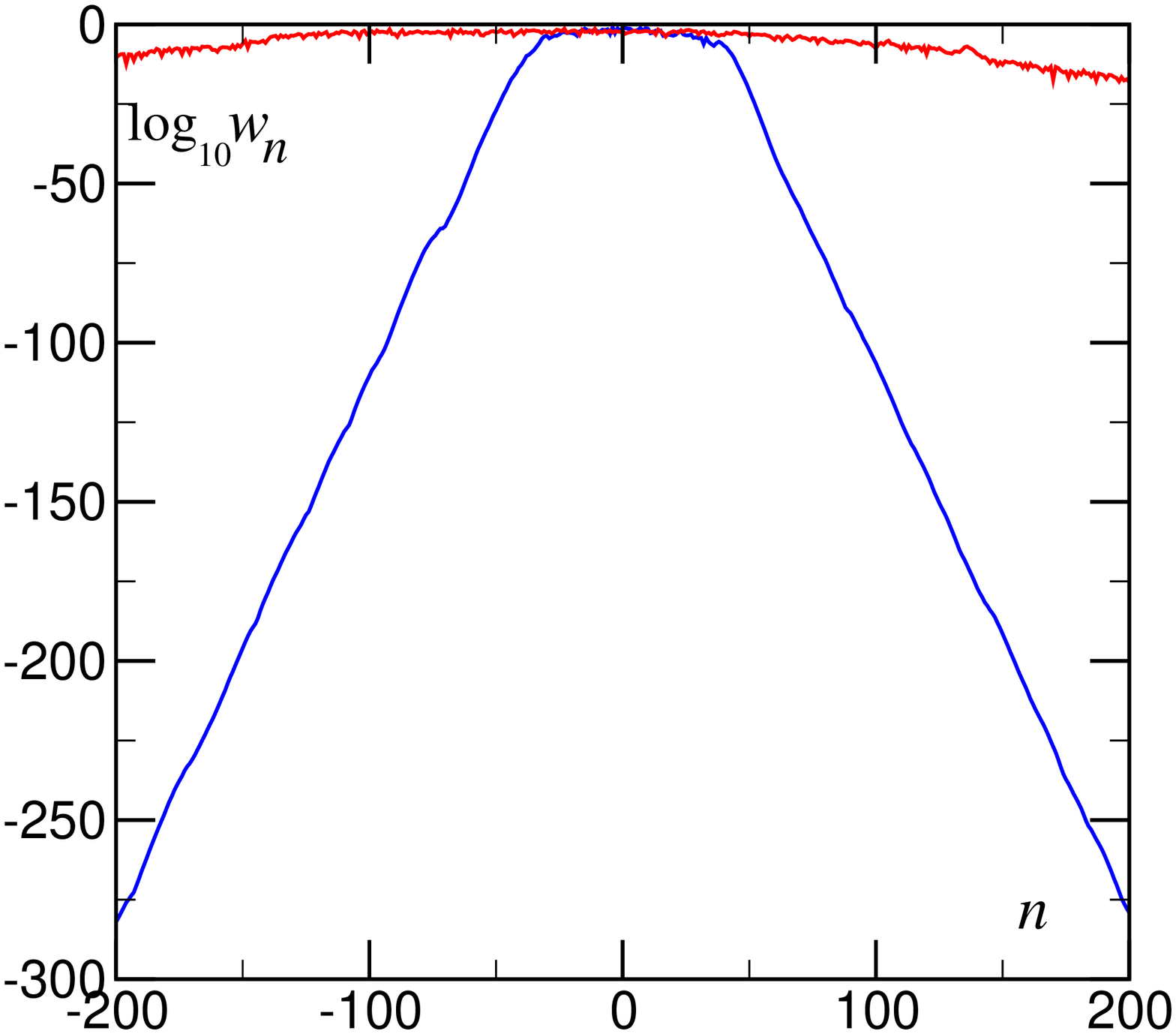}
\hglue 0.5cm
\includegraphics[width=0.22\textwidth]{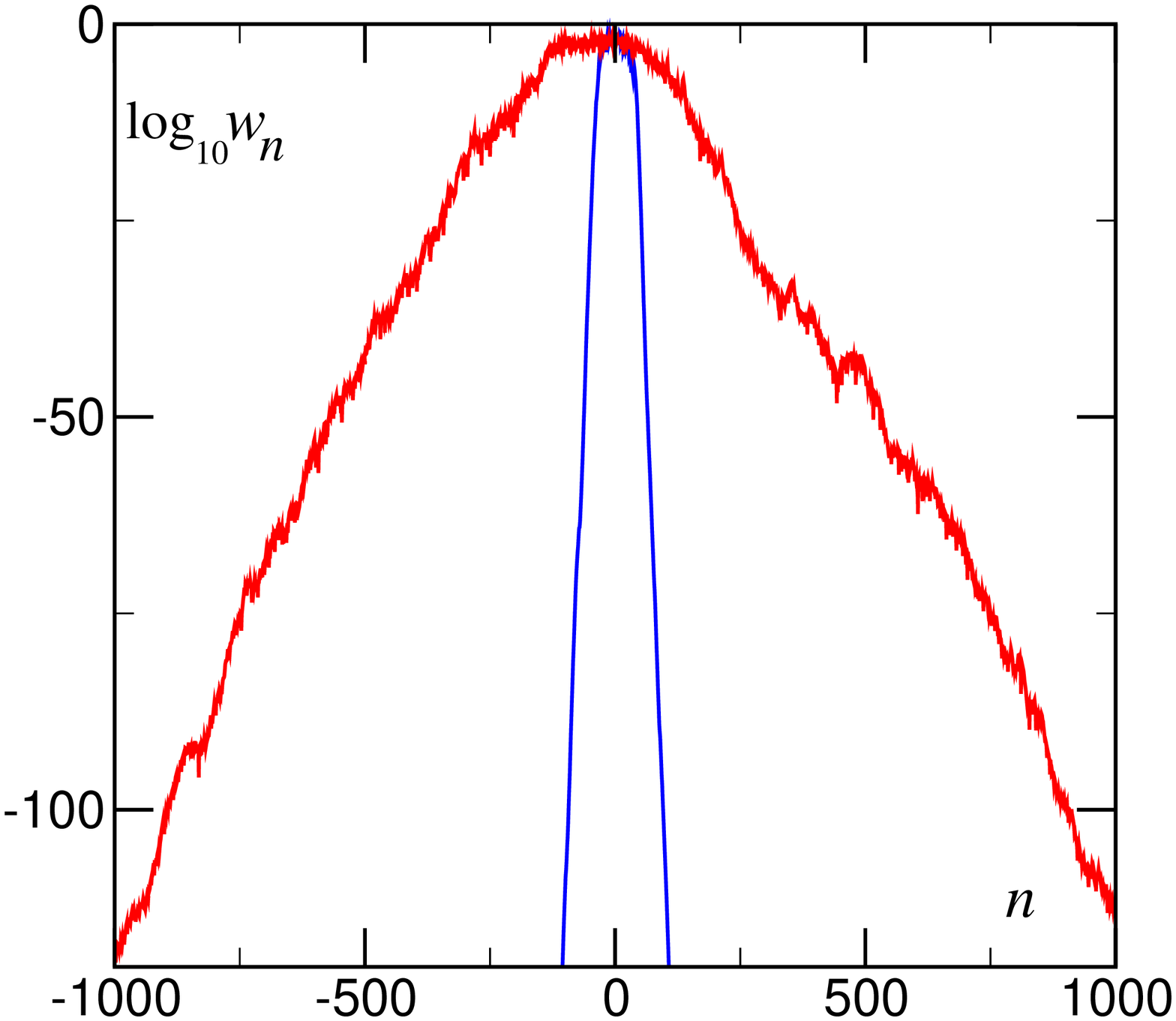}
\end{center}
\vglue 0.2cm
\caption{Probability distribution $w_n$
over linear wave modes $n$ at 
times $10^8$ for the two cases of Fig.~\ref{fig5}
with the same attribution of colors;
two panels show the same data on different scale.
\label{fig6}}
\end{figure}

Let us discuss an example of such a system.
For that we consider the nonlinear Sch\"odinger equation
in a two-dimensional chaotic billiard 
(e.g. the Sinai billiard \cite{sinai}) in a presence
of a monochromatic driving: 
\begin{equation}
i {{\partial {\psi}} / {\partial {\tau}}}
= - \Delta \psi/2 +V(x,y) \psi+ \beta |\psi|^2 \psi
    +F \sin (\omega \tau) x  \psi \; .
\label{eq5}
\end{equation}
Here the potential $V(x,y)$ is determined by
a rigid boundary of the billiard,
$F$ and $\omega$ are the amplitude and 
frequency of a monochromatic force acting
on a particle inside the chaotic billiard.
We assume that the classical dynamics inside the billiard
is chaotic that leads to quantum chaos and ergodicity
of the eigenstates of the linear problem
(see e.g. \cite{billiard}). 
The microwave monochromatic force acting on a 
classical particle creates a diffusive growth of its energy
with time but the quantum interference effects
lead to exponential localization of this 
diffusion  (at $\beta=0$) \cite{prosen},
in a way similar to the KINSE and KNR models at $\beta=0$.
It is interesting to know how the nonlinear term 
with $\beta$ in (\ref{eq5}) will affect this localization and
if nonlinearity can create 
the Kolmogorov flow of energy
from small to large wave vectors.
An analogy with the models of the previous Sections
gives an idea that at weak $F$ and $\beta$ there will be 
no energy flow to high level numbers
of the linear quantum billiard.
However, for the quantum billiard
there is a new element which we discuss below.

Indeed, the density of energy levels
inside the 2D quantum billiard is approximately 
constant, up to quantum fluctuations, \cite{billiard}.
Due to that the energies of quantum  levels
grow with the level number $n$, approximately
as $E_n \sim \rho n$ where $\rho$ is the average density of levels.
Such a behavior of energy levels is different from the
case of DANSE where all unperturbed energy levels are
located inside a finite energy band.
A linear growth of $E_n$ with $n$ corresponds to a presence 
of a static Stark field with an additional term
$\delta E_n =f n$. Such a DANSE model 
with a Stark field has been studied 
in \cite{stark} and it was shown that 
a subdiffusive spreading goes in a way similar to the
DANSE model at moderate values of $f$.
However, for a billiard there is a minimal energy
so that we have only $n \geq 0$
(like a triangular potential). To model such a situation 
one needs to assume that in the DANSE model (\ref{eq3})
there is an additional energy shift 
$\delta E_n = f |n|$. Thus the DANSE model
with such a modulus Stark term
reproduces the energy growth with level number 
typical of the quantum billiard.
We will call this the Stark DANSE model
of billiard (SDANSEBIL model). The numerical simulations
of this SDANSEBIL can be done in the same 
rather efficient way as it is described in \cite{stark}.
The comparison of the results at moderate nonlinearity $\beta=1$
for $f=0$ (DANSE) and $f=0.5$ (SDANSEBIL)
is presented in Fig.~\ref{fig5} and Fig.~\ref{fig6}.
It shows that a finite $f$ reduces the value of $\sigma$
almost by two orders of magnitude and that 
the probability distribution over levels
is localized in a much stronger way compared to the case
with $f=0$. The physical origin of this suppression 
should be attributed to the energy
conservation which for $f>0$
leads to a more rapid decrease of
probability on high levels:
for a homogeneous probability distribution
on levels $0  \leq n \leq n_{max}$
the energy conservation imposes
$|\psi_{n_{max}}|^2 \sim 1/(f n_{max}^2)$
while at $f=0$ the restriction from
norm conservation 
gives slower decay of probability
with $|\psi_{n_{max}}|^2 \sim 1/ n_{max}$.
Due to this reason the spearing over levels is 
suppressed stronger in the presence of finite static field $f$
and triangular form of energies $\delta E_n \propto |n|$.
We note that the case with $\delta E_n \propto n$,
considered in \cite{stark}, does not give
additional restrictions due to cancellations of
negative and positive $n$ contributions.
The obtained results show that the 
transition to ergodicity in the SDANSEBIL model
is practically absent and thus
there is no energy flow to high wave vectors.

Due to similarity between DANSE and KINSE and KNR models discussed
in previous Sections we make a conjecture that
the behavior similar to the one found for
SDANSEBIl model will take place for the evolution
in the nonlinear billiard model (\ref{eq5}).
As a results of that observation it is possible to 
make a conjecture that
there will be no energy flow to high wave vectors
for the Kolmogorov turbulence in Sinai billiard
described by Eq.(\ref{eq5}).
Of course, it would be very interesting to perform
direct numerical simulations of the model (\ref{eq5})
but this would require much more 
advanced and heavy numerical simulations.

\section{Discussion}

In this work we discussed the properties of
weak wave turbulence in finite systems.
On the basis of numerical simulations and analytical
results we argue that the discrete spectrum
of linear frequencies, typical for finite systems,
imposes specific conditions for appearance of 
the Kolmogorov energy flow from 
large to small spacial scales.
In absence of random phase approximation
such a flow can appear only above a chaos
border at a sufficiently large nonlinear coupling
between linear modes and/or strong driving force. 
The considered 
models show that the Anderson localization,
appearing in the wave vector space of linear modes,
can stop the energy flow from
large to small scales if nonlinearity
is below the chaos border and the system is in the regime
of KAM integrability. A similar situation
appears at a small amplitude of energy pumping.
In a qualitative way such a regime 
corresponds to a small wind 
which cannot generate turbulent ocean waves.
Of course, the models analyzed here are relatively
simple and hence, the numerical and experimental studies of 
more realistic systems, like e.g. the model (\ref{eq5}),
are highly desirable. Such studies will allow to understand
a new regime of nonlinear waves where
interplay between Anderson localization, nonlinearity,
KAM integrability and Kolmogorov turbulence in
finite systems opens new interesting and unsolved questions.

Finally, it is useful to note that a quasiperiodic driving
of the KINSE model at $\beta=0$
with two incommensurate frequencies
can create the Anderson transition
with appearance of energy flow to high momentum states
\cite{borgonovi}. Such a transition has been observed recently 
in experiments with cold atoms in kicked optical lattices
\cite{garreau}. Thus it is possible that a transition to
weak turbulence in finite systems is somewhat similar
to the Anderson transition in disordered solids.
In such a scenario the regime of Anderson insulator
corresponds to laminary waves and absence of energy flow from
small wave vectors to large ones,
while the metallic phase allows to have
a turbulent energy flow from small to larger 
wave vectors.
The analysis of possible links
between these phenomena requires further investigations.
The modern techniques of ultra cold atoms and BEC
(see e.g. \cite{raizen,hoogerland},\cite{garreau})
allow to study experimentally
the effects of nonlinear wave interactions in 
such systems.

I thank A.S.Pikovsky for stimulating discussions
and critical remarks.

\end{document}